\documentclass[a4paper,11pt]{article}
\usepackage{jinstpub}
\usepackage{lineno}
\usepackage[version=3]{mhchem}

\usepackage{siunitx}
\DeclareSIUnit\elementarycharge{\text{\ensuremath{e}}}
\usepackage{placeins}
\usepackage{subcaption}

\newcommand{\neqfluence}{$\text{n}_{\text{eq}}/\text{cm}^2$}

\proceeding{24th International Workshop on Radiation Imaging Detectors\\
25-29 JUNE 2023\\
Oslo Science Park}

\title{\boldmath Neutron Radiation induced Effects in 4H-SiC PiN Diodes}

\author[a, 1]{Andreas Gsponer, \note{Corresponding author.}}
\author[a]{Philipp Gaggl,}
\author[a]{Jürgen Maier,}
\author[a]{Richard Thalmeier,}
\author[a]{Simon Emanuel Waid,}
\author[a]{Thomas Bergauer}
\affiliation[a]{Institute of High Energy Physics, Austrian Academy of Sciences, Nikolsdorfer Gasse 12, Vienna}

\emailAdd{andreas.gsponer@oeaw.ac.at}

\abstract{
Silicon carbide (SiC) is a wide band gap semiconductor and an attractive candidate for applications in harsh environments such as space, fusion, or future high luminosity colliders. Due to the large band gap, the leakage currents in SiC devices are extremely small, even after irradiation to very high fluences, enabling operation without cooling and at high temperatures. This study investigates the effect of neutron irradiation on 50$\mu$m p-n 4H-SiC diodes using current-voltage, capacitance-voltage, and charge collection efficiency (CCE) measurements up to neutron fluences of $1\times 10^{16}$ n$_{\text{eq}}$/cm$^2$. The leakage currents of the investigated devices remained extremely small, below 10 pA at 1.1 kV reverse bias.
In the forward bias, a remarkable drop of the current was observed, which was attributed to an increased epi resistivity due to compensation of the epi layer doping by deep-level defects.
The CCE was evaluated for alpha particles from a radioactive source, a 62.4 MeV proton beam at the MedAustron ion therapy center and using UV-TCT.
The charge collection efficiency in reverse bias was shown to scale directly with the 1 MeV equivalent fluence~$\Phi_{\text{eq}}$ as $\text{CCE} \propto \Phi_{\text{eq}}^{-0.63\pm0.01}$.
A CCE better than 50\% was able to be obtained for fluences up to $1 \times 10^{15}$ n$_{\text{eq}}$/cm$^2$.
Because of the low currents in the forward direction, particle detection was also possible in forward bias, where the CCE was found to be increased relative to reverse bias.
Furthermore, a significant dependency on the amount of injected charge was observed, with the CCE surpassing 100\% in alpha and UV-TCT measurements, requiring further systematic investigation.
}

\keywords{Radiation-hard detectors, Radiation damage to detector materials (solid state), Materials for solid-state detectors, Detector design and construction technologies and materials}

\begin{document}
\maketitle
\flushbottom

\section{Introduction}
Silicon Carbide (SiC) detectors have recently become a topic of renewed interest due to the investments in the material by the power electronics industry, specifically large area and high-quality wafers.
The wide band gap of SiC (\SI{3.26}{\electronvolt}) and its high atomic displacement threshold make it an ideal material for applications in harsh environments, such as for fusion reactors, space applications, or at future high luminosity colliders~\cite{De_Napoli_2022}.
The wide band gap leads to extremely low leakage currents, even after irradiation to very high fluences~\cite{rafi_four-quadrant_2018}, which removes the requirement for cooling detectors and allows for operation at high temperatures~\cite{jimenez_high_temperature}.

Radiation damage in SiC has been extensively studied using different devices (p-n junctions, Schottky diodes) and different irradiating particles (protons, electrons, neutrons)~\cite{nava_silicon_2008, rafi_four-quadrant_2018}.
Specifically, the deep-level defects $Z_{1/2}$ and EH$_{6/7}$ have been identified as the major charge carrier traps~\cite{De_Napoli_2022}.
A high concentration of these traps can lead to a compensation of free charge carriers and increased epi layer resistivity as the semiconductor becomes intrinsic~\cite{Kaneko_2011}.
Additionally, the trapping by these defects leads to a decrease in the charge collection efficiency of the device.
Understanding the electrical characteristics and the signal decrease with irradiation is crucial in order to achieve enhanced radiation hardness by techniques such as defect engineering~\cite{Moll_2018}.

In this work, the effect of neutron radiation damage on 4H-SiC p-n diodes is evaluated in an electrical characterization (section~\ref{sec:electrical}) as well as in charge collection efficiency measurements (section~\ref{sec:results:cce})
The work builds on recent investigations using UV-TCT~\cite{gaggl_charge_2022} and alpha particles~\cite{gaggl_performance_2023} using the same devices.
However, new measurements of particle detection in forward bias and data obtained with a \SI{62.4}{\mega\electronvolt} proton beam are presented.
\section{Methods and Materials}
\subsection{Samples}
\begin{figure}[htp]
\centering
\includegraphics[width=.8\textwidth]{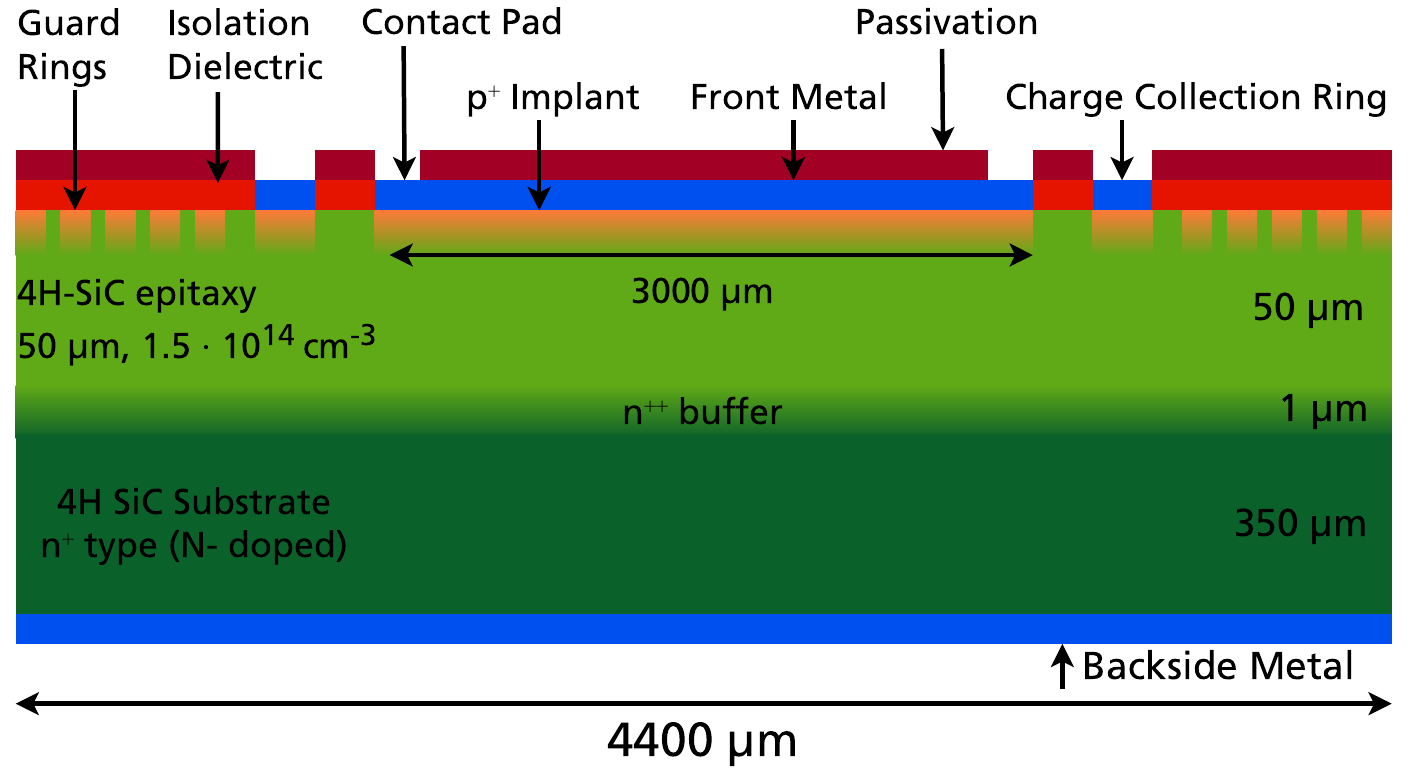}
\caption{Cross-section of 4H-SiC p-n devices manufactured by CNM.\label{fig:crosssection}}
\end{figure}

The investigated samples are 4H-SiC p-n diodes developed and manufactured by IMB-CNM-CSIC.
On top of a \SI{350}{\micro \meter} substrate, an epi layer with a thickness of \SI{50}{\micro \meter} and a doping concentration of \SI{1.5e14}{\per\centi\meter^3} is grown, acting as the active volume.
Figure~\ref{fig:crosssection} shows a cross-section of the device.
Detailed information about the device production can be found in \cite{rafi_four-quadrant_2018}.
The active area of the square samples is $3\times3$\si{\milli\meter^2}, and multiple guard rings are employed in the design to tolerate high bias voltages.
Neutron irradiation of the samples has been carried out at the TRIGA Mark II reactor at the Atominstitut in Vienna, more details can be found in~\cite{gaggl_charge_2022}.

\subsection{Electrical Characterization}
The electrical characterization was performed in a probe station at room temperature (\SI{20}{\celsius}).
In all measurements, the charge collection ring was contacted and connected to ground.
For the I-V measurements, the bias voltage was applied by a Keithley 2657A SMU, while the current was measured by a Keithley 6571b electrometer on the return side of the circuit.
In the C-V measurements, a Keithley 237 SMU was used to apply the bias, and an Agilent 4284A LCR meter was used for the capacitance measurements, with an excitation amplitude of \SI{500}{\milli\volt} and a frequency of \SI{10}{\kilo\hertz}.

\subsection{Charge Collection Efficiency Measurements}
In order to measure the charge deposited by ionizing particles, a Cividec Cx-L shaping charge-sensitive amplifier (CSA) was used.
The CSA has a gain of \SI{12.5}{\milli\volt\per\femto\coulomb}, a shaping time of \SI{1.2}{\micro\second} and a noise specification of \SI{300}{\elementarycharge} + \SI{10}{\elementarycharge \per\pico\farad}.
The CSA features an integrated bias-T, via which a Keithley 2470 SMU applied the bias.
In the alpha measurements, an Eckert \& Ziegler QCRB25 spectrometric mixed nuclide source (\ce{^239Pu}, \ce{^241Am}, \ce{^244Cm}) was used.
The pulse height spectra were acquired using a Rohde \& Schwarz RTO6 oscilloscope in the 16-bit HD mode.
As the source contains three different isotopes, the pulse height spectra were split and analyzed individually per isotope where the energy resolution allowed this.
All measurements were performed in air, and care was taken to maintain a constant source-detector distance of \SI{8}{mm} between the measurements.
\\
The proton beam measurements were performed at MedAustron, an ion therapy facility in Wiener Neustadt, Austria.
MedAustron can provide protons with energies ranging from \num{62.4} to \SI{800}{\mega \electronvolt} with intensities from \si{\kilo\hertz\per\centi\meter^2} to \si{\giga\hertz\per\centi\meter^2}.
The low-intensity beams have been specially commissioned in order to accommodate physics experiments~\cite{Ulrich_Pur_2021}.
In order to maximize the signal inside the detectors and prevent pile-ups, the lowest beam energy of \SI{62.4}{\mega\electronvolt} was chosen, with a medium flux in the sub \si{\mega\hertz\per\centi\meter^2} range.
During data analysis, a Langauss distribution was fitted to the data, and the most probable value of the distribution was extracted.\\
In the UV-TCT setup, a PILAS PIL1-037 laser was used, featuring a wavelength of \SI{370}{\nano\meter} and pulse duration lower than \SI{50}{\pico\second}.
More information about the UV-TCT setup can be found in~\cite{gaggl_charge_2022}.
For some measurements, transient waveforms were obtained using a Cividec C2HV broad-band amplifier with a bandwidth of \SI{2}{\giga\hertz}.
The repetition frequency of the laser was set to \SI{1}{\kilo\hertz} for all measurements, and \num{1000} waveforms were averaged for each data point.

\section{Results}
\subsection{Electrical Characterization}
\label{sec:electrical}
\begin{figure}[htp]
\centering
    \begin{subfigure}[b]{.49\textwidth}
        \includegraphics[height=.66\textwidth]{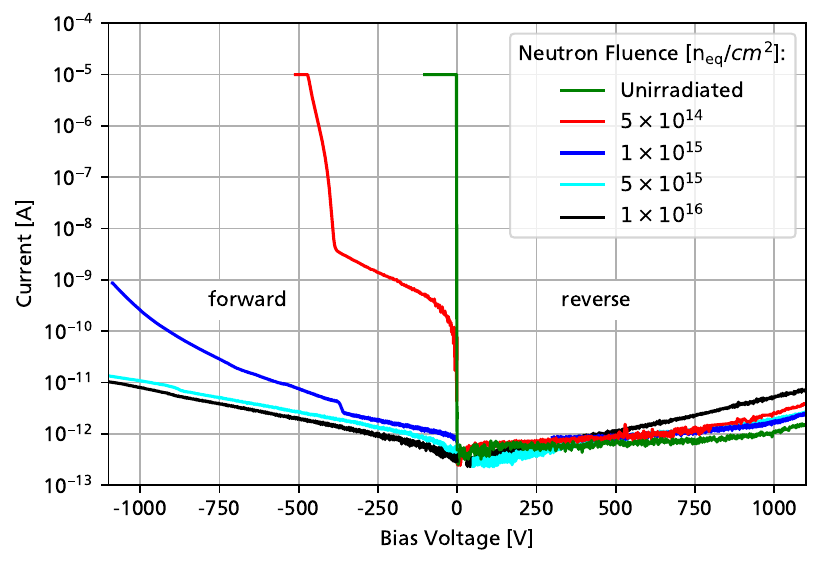}
        \phantomsubcaption
        \label{fig:i-v}
    \end{subfigure}
    \hfill
    \begin{subfigure}[b]{.49\textwidth}
        \centering
        \includegraphics[height=.66\textwidth]{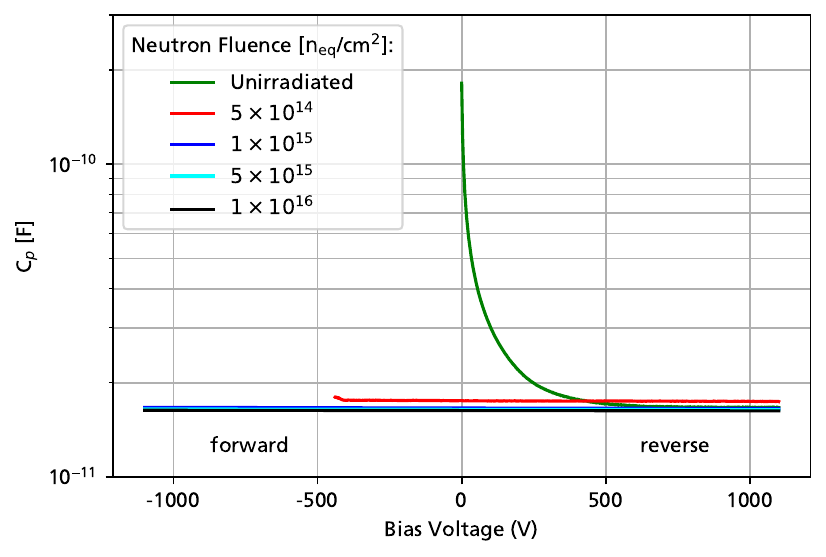}
        \phantomsubcaption
        \label{fig:c-v}
    \end{subfigure}
    \caption{Current-voltage (a) and capacitance-voltage (b) characteristics in forward and reverse bias for different \SI{1}{\mega\electronvolt} equivalent neutron irradiation fluences.\label{fig:electrical}}
\end{figure}

Figure~\ref{fig:i-v} shows the measured I-V characteristics in forward and reverse bias.
In reverse bias, the leakage currents at room temperature remain extremely low compared to silicon devices~\cite{Moll_2018}, below \SI{10}{\pico\ampere}, even after irradiation to fluences as high as $10^{16}$ \neqfluence.
The low currents be attributed to the high atomic displacement threshold of SiC and the low charge carrier generation by the induced deep-level defects~\cite{Rafi_2023}.
In forward bias, a drastic decrease of the current is observed.
Already at a fluence of $1\times 10^{15}$ \neqfluence, the forward current is reduced below \SI{1}{\nano\ampere} at \SI{1000}{\volt} bias.
The forward current is identical to the reverse current for the highest irradiation fluences.
The decrease in forward current can be explained by deep-level defects ($Z_{1/2}$ and EH$_{6/7}$), which trap the charge carriers introduced by the epi-doping, increasing the resistivity as the semiconductor becomes intrinsic~~\cite{Kaneko_2011}.
Additionally, the defect-induced reduction of charge carrier lifetime results in a loss of the current modulation properties of the diode~\cite{Kaji_2015}.
\
Figure~\ref{fig:c-v} shows the capacitance-voltage \mbox{(C-V)} measurements for unirradiated and irradiated samples.
A diode-like depletion is observed for the unirradiated sample, with an estimated full depletion voltage of \SI{325}{\volt}.
For the irradiated samples, the capacitance can also be measured in forward bias, due to the low leakage currents discussed previously.
The measured capacitance is constant with the bias voltage, aside from the sample irradiated to $5\times10^{14} \text{n}_{\text{eq}}$, which shows a minor distortion around \SI{-450}{\volt} due to the large currents present.
This constant capacity is in accordance with the \SI{50}{\micro\meter} thickness of the epi-layer.
This, again is compatible with the epi layer becoming intrinsic due to the doping compensation by acceptor-type defects and has been observed previously for similar 4H-SiC devices~\cite{Moscatelli_2006, rafi_four-quadrant_2018, Kaneko_2011}.
Due to the flat C-V curves, no carrier removal rates could be extracted from the data, for which lower irradiation fluences should be investigated.
\subsection{Charge Collection Efficency}
\label{sec:results:cce}
Due to the low currents in forward bias, discussed in section \ref{sec:electrical}, signals from ionizing particles could also be obtained in forward bias.
This has been observed for silicon devices as well, where the charge collection efficiency (CCE) increased in forward bias.
This increase in CCE was attributed to an enhanced electric field uniformity and trapping filling by the forward current of the diode~\cite{forward_Lutz_1996, forward_Beattie_2000}.
Recent measurements on similar 4H-SiC devices using alpha particles did not show an increase in the CCE in forward bias, but a better energy resolution, especially at low bias voltages~\cite{Rafi_2023}.
This section presents CCE results in reverse and forward bias for alpha particles from a radioactive source, a \SI{62.4}{\mega\electronvolt} proton beam, and UV-TCT.
\subsubsection{Alpha Particles}
\label{sec:results:alpha}
\begin{figure}[htp]
\centering
    \includegraphics[height=.25\textheight]{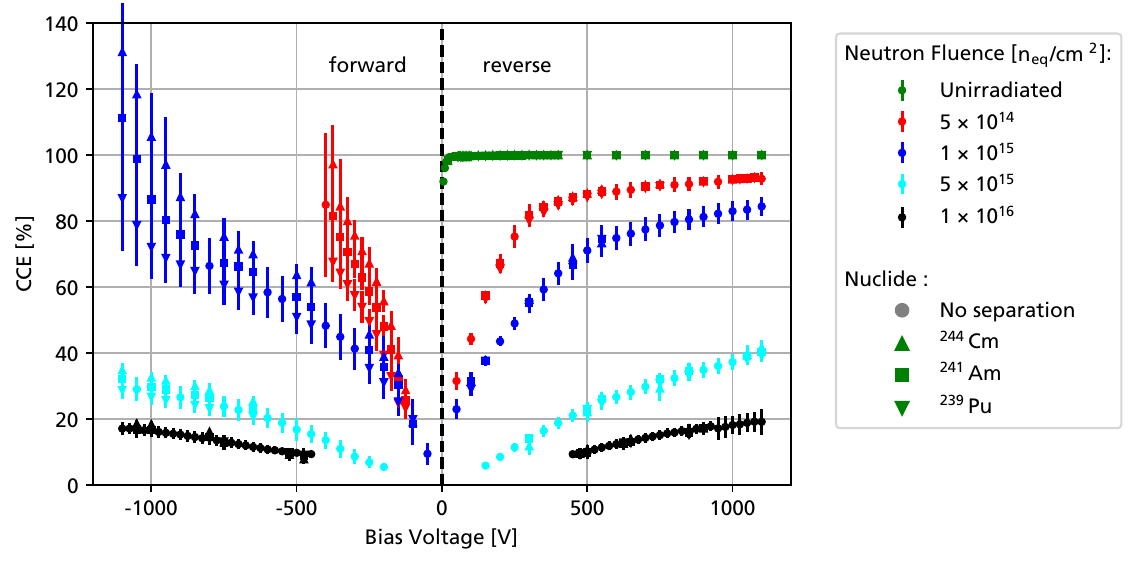}
    \caption{Charge collection for alpha particles in forward and reverse bias. The CCE has been normalized to each of the three isotopes present in the source where possible.}
    \label{fig:cce:alpha}
\end{figure}

Figure~\ref{fig:cce:alpha} shows the measured charge collection efficiency as a function of forward and reverse bias for different neutron fluences.
In reverse bias, charge collection efficiencies above \SI{80}{\percent} can be obtained for fluences up to $1 \times 10^{15}$ \neqfluence.
This is slightly higher than previously published results for the same samples~\cite{gaggl_performance_2023} and can be attributed to the usage of a charge-sensitive amplifier (CSA) instead of a broad-band amplifier.
After normalization to the decay energy, no difference can be observed in the reverse bias CCE for the different isotopes.\\
In forward bias and for the lowest neutron fluence, $5 \times 10^{14}$ \neqfluence, the CCE is first lower than in reverse bias but surpasses it at \SI{-400}{\volt}.
At the highest forward bias voltages, the shot noise due to the larger forward current (see section~\ref{sec:electrical}) obscures the signal.
A similar behavior can be observed for the next fluence, $1 \times 10^{15}$ \neqfluence.
Here, the charge collection even surpasses the non-irradiated device (\SI{100}{\percent}).
Additionally, the CCE is not uniform for the different isotopes, with an energy dependency that is not observed in reverse bias.
At the two highest irradiation fluences, the CCE of the devices is very similar in forward and reverse bias.
As previously discussed, increased CCE in forward bias can be attributed to a higher electric field uniformity and trap filling.
However, this cannot explain a charge collection efficiency above \SI{100}{\percent}.
Possible explanations are discussed in the context of UV-TCT results in section~\ref{sec:results:tct}.
\FloatBarrier
\subsubsection{Proton Beam}
\label{sec:results:proton}
\begin{figure}[htp]    \centering
    \includegraphics[height=.25\textheight]{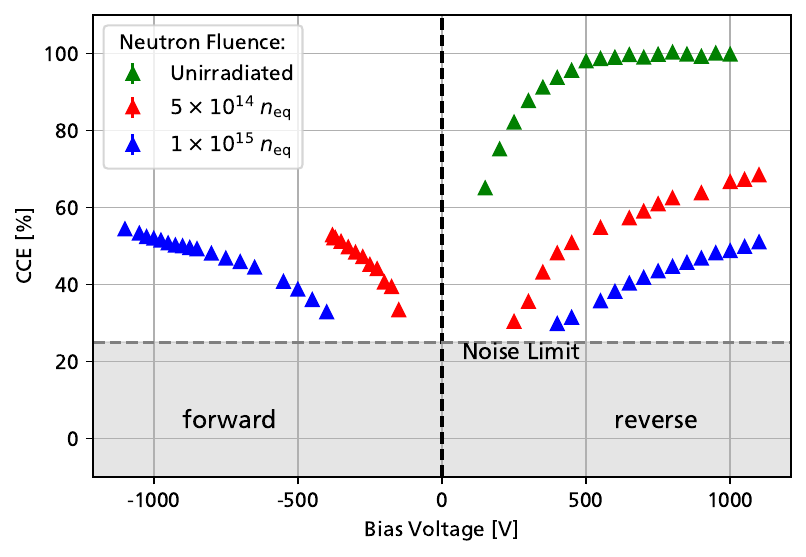}
    \caption{Charge collection efficiency for \SI{62.4}{\mega\electronvolt} protons at MedAustron.\label{fig:cce:proton}}
\end{figure}

In measurements with a proton beam, only data for the two lowest irradiation fluences could be obtained, see figure~\ref{fig:cce:proton}.
This is mainly due to the limited charge deposition in the thin detector (\SI{50}{\micro\meter}) epi and a non-ideal RF shielding of the detector, which limited the signal-to-noise ratio.
However, at the highest bias, charge collection efficiencies better than \SI{50}{\percent} were obtained below a fluence of $1\times10^{15}$\neqfluence, in both forward and reverse bias.
For the same bias voltage, the charge collection efficiency is improved in forward bias over reverse bias.

\FloatBarrier
\subsubsection{UV-TCT}
\label{sec:results:tct}
\begin{figure}[htp]
    \centering
    \begin{subfigure}[b]{.49\textwidth}
        \centering
        \includegraphics[height=.65\textwidth]{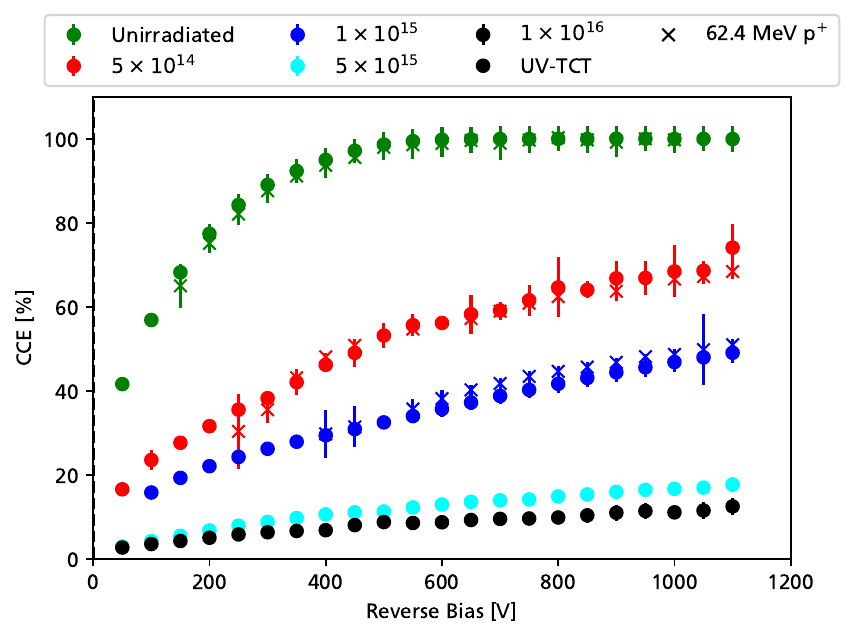}
        \phantomsubcaption
        \label{fig:cce:tct:reverse_vs_voltage}
    \end{subfigure}
    \hfill
    \begin{subfigure}[b]{.49\textwidth}
        \centering
        \includegraphics[height=.67\textwidth]{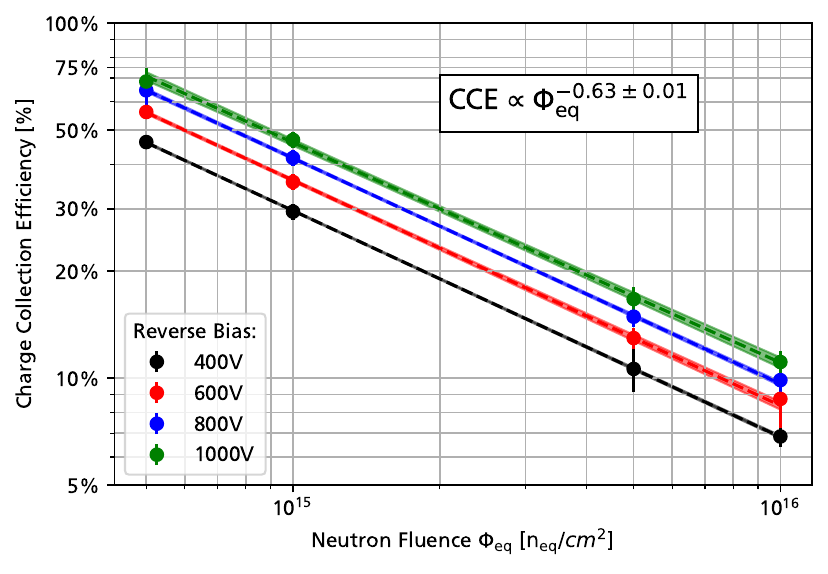}
        \phantomsubcaption
        \label{fig:cce:tct:reverse_vs_fluence}
    \end{subfigure}
    \caption{(a): Charge collection efficiency versus reverse bias voltage for UV-TCT and proton beam data. (b): Power-law fit of CCE to the \SI{1}{\mega\electronvolt} neutron equivalent irradiation fluence at fixed reverse bias voltages. \label{fig:cce:tct:reverse}}
\end{figure}

In the UV-TCT measurements, CCEs as low as \SI{5}{\percent} were able to be quantified due to the large amount of charge injected~\cite{gaggl_charge_2022} and waveform averaging.
For reverse bias, the CCE depicted in Figure~\ref{fig:cce:tct:reverse_vs_voltage} agrees very well with the proton beam measurements and previous investigations~\cite{gaggl_charge_2022}.
Above the full depletion voltage, the CCE decrease irradiation can be very well described by a power law relation to the \SI{1}{\mega\electronvolt} neutron equivalent fluence $\Phi_{\text{eq}}$, similar to silicon devices~\cite{Moll_2018}.
This power law is very well reproduced for different reverse bias voltages, and a dependency of $\text{CCE} \propto \Phi_{\text{eq}}^{-0.63\pm0.01}$ was obtained, see Figure~\ref{fig:cce:tct:reverse_vs_fluence}.

\begin{figure}[htp]
    \centering
    \begin{subfigure}[t]{.49\textwidth}
        \centering
        \includegraphics[height=.64\textwidth]{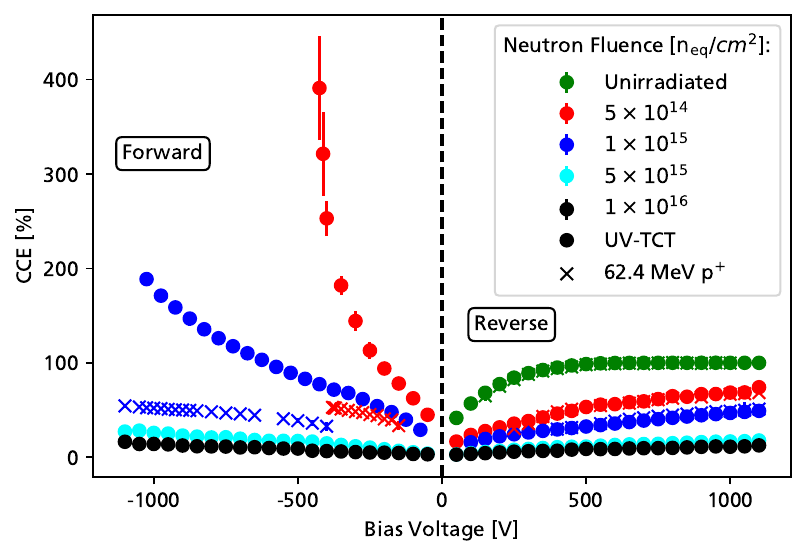}
        \phantomsubcaption
        \label{fig:cce:tct:forward:forward_and_reverse}
    \end{subfigure}
    \hfill
    \begin{subfigure}[t]{.49\textwidth}
        \centering
        \phantomsubcaption
        \includegraphics[height=.64\textwidth]{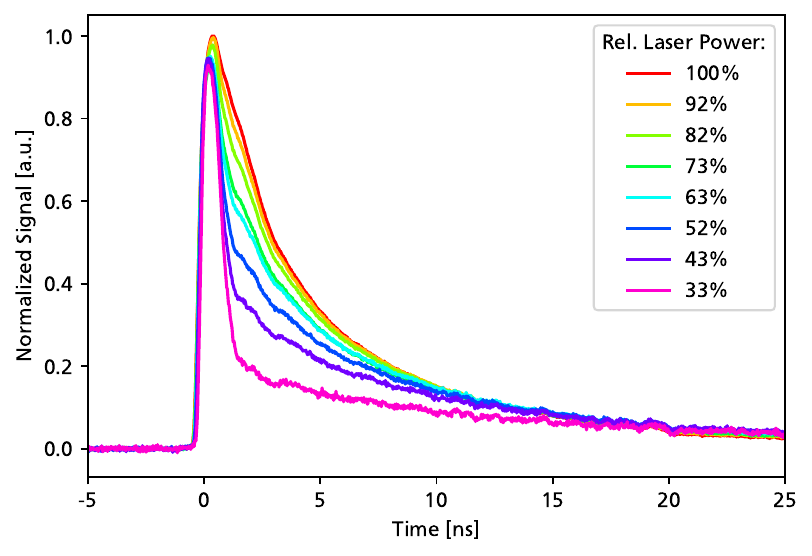}
        \label{fig:cce:tct:forward:transients}
    \end{subfigure}
    \caption{(a): Charge collection efficiency (CCE) in forward and reverse bias for UV-TCT and proton beam data. (b): Normalized waveforms at \SI{-450}{\volt} (forward bias) for the sample irradiated to $5\times10^{14}$ \neqfluence~, measured using using a broad-band amplifier and different laser power levels. \label{fig:cce:tct:forward}}
\end{figure}

A significant discrepancy can be observed in the forward bias, depicted in Figure~\ref{fig:cce:tct:forward:forward_and_reverse}.
An increase in the charge collection efficiency above \SI{100}{\percent} was observed for the two lowest fluences, with an exponential behavior on the bias voltage.
The increased CCE is similar to the results obtained using alpha particles in section~\ref{sec:results:alpha}.
Furthermore, it was observed experimentally that the CCE depends on the charge injected by the laser.
Figure~\ref{fig:cce:tct:forward:transients} shows transient waveforms obtained using a broad-band amplifier at different charge injection levels.
When normalizing the transient signals to the injected charge, the amplitude of the transients remains quite similar.
However, there is a long (> \SI{10}{\nano\second}) tail in the signal, which contributes significantly to the integrated signal and is a function of the injected charge.
This effect has also been observed recently in TPA-TCT measurements~\cite{alvarez_TPA_TCT}.
More work is required to understand this effect and to investigate systematically the dependency on the amount of injected charge.
Studies with different laser repetition frequencies or a continuous light illumination (see~\cite{Mandi_2004}) could explain if this effect is related to trap filling and modulation of the forward current by injected charge carriers.

\section{Conclusions and Outlook}
Neutron-irradiated 4H-SiC p-n diode detectors have been investigated electrically, using ionizing particles and using UV-TCT.
Even for the highest investigated neutron fluence, $1 \times 10^{16}$ \neqfluence, the leakage currents remain extremely low, below \SI{10}{\pico \ampere} at \SI{1.1}{\kilo\volt} bias.
A reduction in the forward current due to radiation damage has been observed, which can be attributed to compensation by deep-level defects leading to an intrinsic (semi-insulating) epi-layer~\cite{Kaneko_2011}.
The charge collection efficiency (CCE) after irradiation has been studied for ionizing particles and using a UV laser.
In reverse bias, a power law dependency of the CCE on the neutron irradiation fluence was observed as $CCE \propto \Phi_{\text{eq}}^{-0.63\pm0.01}$.
This power law also holds for different bias voltages above \SI{400}{\volt}.\\
The low currents in forward bias allow the detectors to be used in forward bias for particle detection as well.
In forward bias, the CCE was shown to be improved compared to the reverse bias, as has been observed for silicon devices~\cite{forward_Lutz_1996, forward_Beattie_2000}.
At the lowest irradiation fluences, a dependency of the CCE on the amount of injected charge was observed, and CCEs above \SI{100}{\percent} were obtained at very large charge injections.
This effect needs to be studied more closely to be understood, with studies varying the amount of injected charge or by injecting free charge carriers using UV-light illumination.
Furthermore, measurements at different temperatures, which affect the current in forward bias~\cite{Rafi_2023}, could help to study the dependency of the CCE on the forward current.
Finally, device simulation in TCAD software could be a very valuable tool for understanding the electric fields and the charge collection in devices after deep-level defects have been introduced by radiation damage.

\acknowledgments
This project has received funding from the Austrian Research Promotion Agency FFG, grant number 883652. Production and development of the 4H-SiC samples was supported by the Spanish State Research Agency (AEI) and the European Regional Development Fund (ERDF), ref. RTC-2017-6369-3.

\bibliographystyle{JHEP}
\bibliography{biblio.bib}

\end{document}